\documentclass[10pt,conference,letterpaper]{IEEEtran}
\usepackage{cite}
\usepackage{amsthm,amsmath,amssymb,amsfonts}
\usepackage{graphicx}
\usepackage{textcomp}
\usepackage{xcolor}
\usepackage{subfigure}
\usepackage{algorithm}
\usepackage{algorithmic}
\usepackage{diagbox}
\usepackage{makecell}
\usepackage{mathrsfs}

\begin{document}
\title{Hierarchical Reinforcement Learning with Opponent Modeling for Distributed Multi-agent Cooperation}
\author{Zhixuan Liang, Jiannong Cao, \textit{Fellow, IEEE}, Shan Jiang\textsuperscript{\textsection}, Divya Saxena, Huafeng Xu\\
Department of Computing, The Hong Kong Polytechnic University
}

\maketitle
\begingroup\renewcommand\thefootnote{\textsection}
\footnotetext{Corresponding author: Shan Jiang}
\endgroup

\begin{abstract}
Many real-world applications can be formulated as multi-agent cooperation problems, such as network packet routing and coordination of autonomous vehicles. The emergence of deep reinforcement learning (DRL) provides a promising approach for multi-agent cooperation through the interaction of the agents and environments. However, traditional DRL solutions suffer from the high dimensions of multiple agents with continuous action space during policy search. Besides, the dynamicity of agents' policies makes the training non-stationary. To tackle the issues, we propose a hierarchical reinforcement learning approach with high-level decision-making and low-level individual control for efficient policy search. In particular, the cooperation of multiple agents can be learned in high-level discrete action space efficiently. At the same time, the low-level individual control can be reduced to single-agent reinforcement learning. In addition to hierarchical reinforcement learning, we propose an opponent modeling network to model other agents' policies during the learning process.
In contrast to end-to-end DRL approaches, our approach reduces the learning complexity by decomposing the overall task into sub-tasks in a hierarchical way. To evaluate the efficiency of our approach, we conduct a real-world case study in the cooperative lane change scenario. Both simulation and real-world experiments show the superiority of our approach in the collision rate and convergence speed.
\end{abstract}


\section{Introduction}
Many complex real-world applications can be modeled as multi-agent cooperation, such as network packet routing \cite{silva2019reinforcement}, energy distribution \cite{zhao2012energy}, and coordination of autonomous vehicles \cite{jiang2016ensemble}. In these applications, the agents need to make individual decisions considering the mutual influence. Traditional solutions of multi-agent cooperation are mostly model-based, in which the agents and their interactions are modeled using physical formulas and prior knowledge \cite{ma2016event}\cite{sahni2019middleware}. These approaches fail to adapt to dynamic, stochastic, and complex environments. Recently, the development of deep reinforcement learning (DRL) \cite{wang2019pattern} provides a promising solution through a \textit{trial-and-error} process \cite{hernandez2019survey}. At each step, the agent observes the environment, selects the optimal action, and receives rewards as feedback signals related to the team performance. The goal of each agent is to learn the policies maximizing the accumulated rewards received from the environment.

\begin{figure}[t]
\centering
\includegraphics[width=\linewidth]{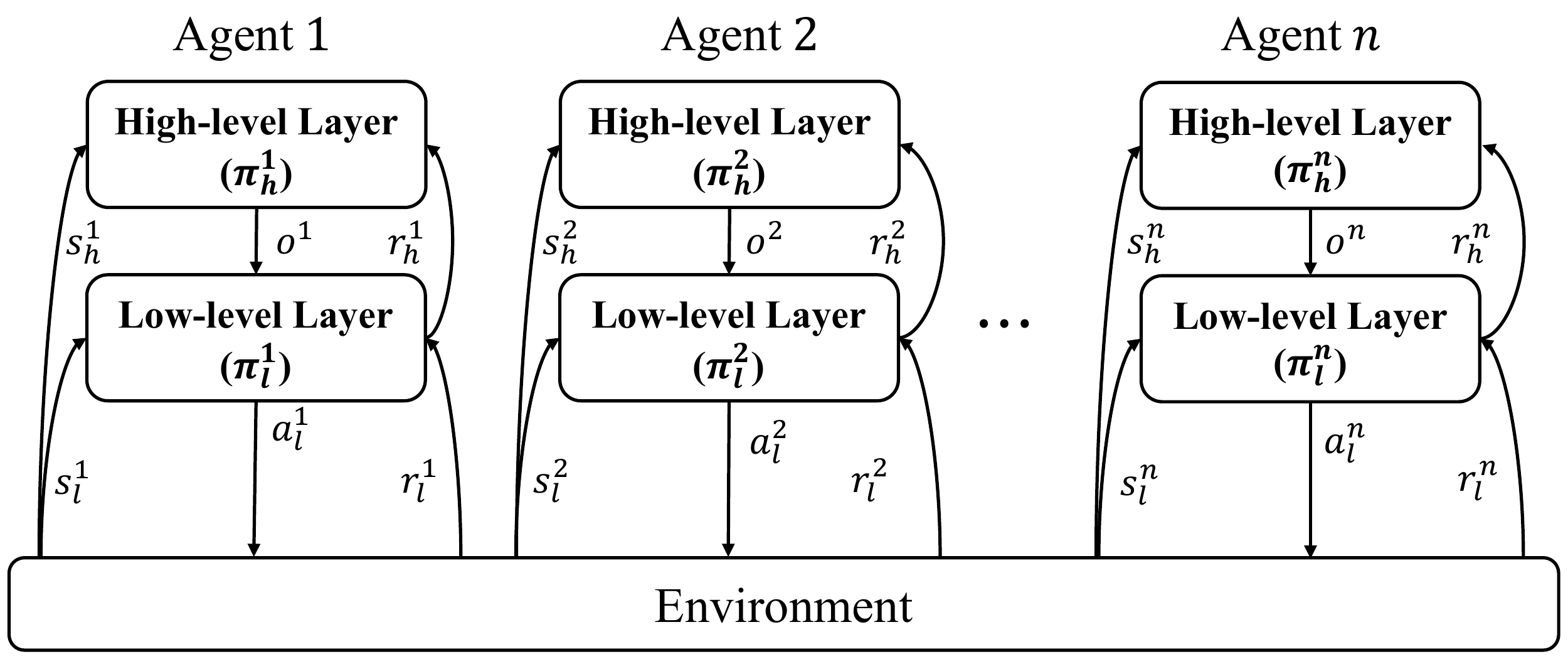}
\caption{Illustration of hierarchical reinforcement learning for distributed multi-agent cooperation. Each agent maintains a high-level cooperation layer and a low-level individual control layer.}\label{Fig:HMARL}
\end{figure}

Existing DRL approaches for multi-agent cooperation can be classified into three categories: centralized reinforcement learning, centralized training with decentralized execution, and distributed reinforcement learning. Centralized reinforcement learning aims to train a centralized value function that selects actions for all the agents. Such a method is hard to be extended to large-scale scenarios due to the exponential action spaces when the number of agents increases. An alternative approach is centralized training with decentralized execution (CTDE), which trains a critic network to estimate state-action pairs' values. At the same time, each agent maintains an individual actor network for decentralized action selection. By training such a critic network considering the states and actions of other agents, all the agents can learn to cooperate in certain states. Some previous works have applied CTDE for real-world applications, such as flocking \cite{zhu2020multi} and pathfinding \cite{damani2021primal}. However, the number of features in the critic network needs to be scaled up linearly (in the best case) or exponentially (in the worse case) as the number of agents increases. Besides, the gain of learning a centralized critic is likely to be minimal in the sparse interaction scenarios \cite{lyu2021contrasting}.

Recent developments in DRL address the above limitations through decentralized training with decentralized execution (DTDE), where each agent acts and learns to cooperate in a distributed manner\cite{zhang2018fully}\cite{iqbal2019actor}\cite{qu2019value}. However, it is non-trivial to learn in a distributed way because the training is non-stationary, and it is hard to represent individual agents' policies. In particular, standard DRL approaches relying on end-to-end models suffer from the curse of dimensionality of multiple agents in continuous action spaces. It demands large-scale models and a long time for training to learn a mapping function from the state to the continuous action space. Furthermore, the independence of individual policy learning leads to poor coordination. Finally, the dynamicity of the agents' policies makes it nearly impossible to employ experience replay \cite{schaul2015prioritized} that is crucial for stabilizing DRL.

In this paper, we propose a novel hierarchical deep reinforcement learning approach for distributed multi-agent cooperation. First, we decompose the policy space of each agent into a high-level cooperation layer and a low-level individual control layer, which is shown in Fig.~\ref{Fig:HMARL}. The cooperation of multiple agents can be efficiently learned in high-level discrete action space, while the low-level individual control can be reduced to independent reinforcement learning. Besides, we introduce an opponent modeling mechanism to model other agents' high-level decisions. The learned opponent models will encourage cooperation behaviors and stabilize DRL. In contrast to the standard end-to-end DRL model, we reduce the learning complexity by decomposing the overall task into sub-tasks that are easier to solve hierarchically.
Furthermore, we conduct a case study on cooperative lane change and present the hierarchical decision-making model with the specific design of states, actions, and rewards. Finally, we evaluate our approach in a simulation environment and a real-world testbed. The main contributions of this paper are as follows:
\begin{itemize}
    \item We study the problem of distributed multi-agent cooperation in continuous action space and propose a hierarchical deep reinforcement learning approach, which decomposes the overall task into high-level option selection and low-level individual control.
    \item To address the non-stationary training issue in distributed multi-agent learning, we introduce an opponent modeling mechanism to model other agents' high-level policies.
    \item We conduct extensive experiments on cooperative lane change and compare our approach with state-of-art baselines in simulation and a real-world testbed. The experimental results show the superiority of our approach in both task performance and convergence speed.
\end{itemize}

The rest of the paper is organized as follows. Sec.~\ref{sec:related-work} summarizes the previous works of multi-agent reinforcement learning and hierarchical reinforcement learning. Sec.~\ref{sec:preliminaries} introduces the notations and preliminaries of deep reinforcement learning. Sec.~\ref{sec:proposed-method} presents a novel hierarchical reinforcement learning algorithm with an opponent modeling mechanism. Sec.~\ref{sec:case-study} showcases a real-world case study of hierarchical reinforcement learning on the multi-vehicle cooperative lance change scenario. Sec.~\ref{sec:exp} shows the experimental results in simulation and a real-world bed. Finally, Sec.~\ref{sec:dis-con} concludes the paper and discusses the future directions.

\section{Preliminaries}\label{sec:preliminaries}
In this section, we introduce the definitions of Markov decision process (MDP), reinforcement learning (RL), and hierarchical reinforcement learning (HRL).

\subsection{Markov Decision Process}
Many decision-making problems can be mathematically modeled as a Markov decision process (MDP). An MDP is defined as a four-tuple $(S,A,r,T)$, in which $S$ is the set of states, $A$ is the set of available actions, $r: S\times A \times S \rightarrow R$ is the reward function, and $\mathcal{T}:S\times A\times S\times \rightarrow [0,1]$ is the probability function of state-action-state transition. A stochastic policy function is defined as $\pi: S \times A \rightarrow [0,1]$ and a deterministic policy function  $\mu: S \rightarrow R^{|A|}$ is defined to select an action for current state. The goal of solving MDP is to find a policy which selects actions maximizing the cumulative rewards $R=\sum_{t=0}^{N} r_t$.

Given a policy $\pi$ and a state $s$, the value function $V: S \rightarrow R$ calculates the expected sum of discounted rewards using the following formula:
\begin{align*}
V^{\pi}(s)=\mathbb{E}_{\pi}\left[\sum\nolimits_{t=0}^{N} \gamma^{t} r_{t} \mid s\right]
\end{align*}
where $\gamma$ is the discount factor and $r_t$ is the reward received at each time step. In addition, given a policy $\pi$, a state $s$, and an action $a$, the state-action value function $Q: S \times A \rightarrow R$ calculates the expected sum of discounted rewards from a given state as follows:
\begin{align*}
Q^{\pi}(s,a)=\sum\nolimits_{\pi}\left[\sum\nolimits_{t=0}^{N} \gamma^{t} r_{t} \mid s,a\right]
\end{align*}

Solving an MDP requires computing the state-value function and state-action value function. The reward function and state transition are entirely known, these functions can be solved by iterating the Bellman equations, and the decision-making problem can be solved by using dynamic programming \cite{sutton2000policy}. However, the reward function and state transition are unknown to agents in most cases.

\subsection{Reinforcement learning}
Reinforcement learning provides multiple methods for solving MDPs, classified as value-based and policy-based methods.

In value-based method, the state-action-value function (or state-value function) is estimated using temporal different (TD) methods as follows:
\begin{align*}
\begin{split}
& Q(s_t,a_t) \\ =  & Q(s_t,a_t) +\alpha(r_t+\gamma \max_{a^{t+1}} Q(s^{t+1}, a^{t+1})-Q(s_t,a_t))
\end{split}
\end{align*}
Then the greedy policy can be derived by selecting the action with highest Q-value at each time step as follows:
\begin{align*}
    \pi(s_t)=\arg \max_a Q(s_t,a_t)
\end{align*}
Such a method is referred to as the Q-learning \cite{busoniu2008comprehensive}.

Traditional tabular Q-learning suffers from the issue of high dimensionality in state space. Deep Q-learning (DQN) \cite{mnih2015human} addresses the issue by employing a deep neural network to approximate the Q-function. Besides, DQN is a class of off-policy reinforcement learning methods that uses a \textit{replay memory} to store the transition four-tuple $(s, a, r, s')$ and these data can be sampled to train the Q-network. The loss function of training Q-network is as follows:
\begin{align*}
    \mathcal{L}(\theta)=\mathbb{E}_{s, a, r, s_{'}^{\prime}}\left[\left(Q\left(s_{},{a}_{};\theta\right)-y\right)^{2}\right]
\end{align*}
where $y=r+\gamma \max _{{a} \in \mathcal{A}}\left(Q\left({s}_{'}, {a};\theta^{-}\right)\right)$ and $\gamma \in [0,1)$ is the discount factor. In $y$, the variable $\theta^{-}$ is the parameter of the \textit{target network} that is periodically copied from $\theta$ that is kept constant for a number of iterations.

An alternative model-free approach is the policy-based method, which directly learns a policy $\pi$ parameterized by $\theta^{\pi}$. The objective of these approaches is to adjust the parameters $\theta^{\pi}$ in order to maximize the function $J(\theta)=\mathbb{E}_{s \sim p^{\pi}, a \sim \pi_{\theta}}[R]$, where $R$ denotes the expected accumulative rewards which are usually approximated by $Q(s,a)$. In the actor-critic theory, the policy network $\pi_{\theta}$ can be considered an actor network, and the state-action value function $Q(s, a)$ can be considered a critic network. According to the policy gradient theorem \cite{sutton2000policy}, the gradient of this objective function is defined as follows:
\begin{align*}
\nabla_{\theta} J\left(\pi_{\theta}\right)=\mathbb{E}_{s \sim \rho^{\pi}, a \sim \pi_{\theta}}\left[\nabla_{\theta} \log \pi_{\theta}(a | s) Q^{\pi}(s, a)\right]
\end{align*}
where $\rho^{\pi}$ is the state distribution under policy $\pi_{\theta}$ and $\pi_{\theta}(a|s)$ is the probability of select the action $a$ under the state $s$. At each time step, the agent samples an action from the distribution generated from the policy network.

Deep deterministic policy gradient (DDPG) is an actor-critic algorithm extended from stochastic policy gradient to deterministic policy gradient \cite{silver2014deterministic}\cite{lillicrap2015continuous}. It employs a deep neural network parameterized by $\theta^\mu$ to approximate the deterministic policy $\mu_{\theta}:S \rightarrow A$ and a neural network parameterized by $\theta^{Q}$ to approximate the action-value function $Q(s,a|\theta^{Q})$. The critic network is learned using the Bellman equation as in Q-learning, and the actor network is updated by applying the chain rule to the objective function $J$: 
\begin{align*}
    \nabla_{\theta} J(\theta)=\mathbb{E}_{s \sim \mathcal{D}}\left[\left.\nabla_{\theta} \mu_{\theta}(a \mid s) \nabla_{a} Q^{\mu}(s, a)\right|_{a=\mu_{\theta}(s)}\right]
\end{align*}

\subsection{Hierarchical Reinforcement learning with Temporal Abstraction}
Human decision-making often involves choosing among temporally extended courses of action over a broad range of time scales \cite{pateria2021hierarchical}. Learning and operating over different levels of temporal abstraction is a critical challenge in tasks involving long-term planning.
Sutton et al. \cite{sutton1999between} extended reinforcement learning framework to include temporally abstract actions, representations that group together a set of interrelated actions (for example, moving to block $A$, driving another lane, passing the ball to another person).
These representations can be translated into a series of individual actions, which are described as \textit{temporal abstraction}. A recent extension of this direction is hierarchical deep reinforcement learning with temporal abstraction proposed by Kulkarni et al. \cite{kulkarni2016hierarchical}. In their approach, the agent uses a two-level decision-making model, i.e., a \textit{meta-controller} and a \textit{controller}. The meta-controller receives a state $s_t$ and chooses a goal $g_t\in \mathcal{G}$, where $\mathcal{G}$ denotes the set of all possible current goals. The $g_t$ will remain for $T$ time steps, or the terminated state is reached. The Q-value function for the controller is:
\begin{align*}
    Q_l(s,a;g)=\mathrm{E}\left[r_{l}+\gamma \max _{a_{'}} Q_{l}\left(s_{'}, a_{'} ; g\right) \right]
\end{align*}
where $g$ is the goal and $r_{l}$ is low-level reward given by a goal-driven reward function $R(s_t,a_t,g_t)$. This low-level reward is related to performing the goal (temporal abstraction), which is also regards as \textit{intrinsic reward}. The design of intrinsic reward and termination is still an open question in hierarchical reinforcement learning.

Similarly, the Q-value function of the meta-controller is:
\begin{align*}
    Q(s,g)=\mathrm{E}[r_{h}+\gamma \max_{g^{'}} Q(s^{'},g^{'})]
\end{align*}
where $r_{h}=\sum_{t=0}^{T}r$ is the accumulative reward received from the environment during the $T$ time steps.

\section{HERO: Hierarchical Reinforcement Learning with Opponent Modeling} \label{sec:proposed-method}
This section introduces the sequential decision-making problem in distributed multi-agent systems. Then, we propose HERO, a general hierarchical decision-making model with high-level cooperative decision-making and low-level individual control. Note that, in many applications, the primitive actions of agents are regarded as individual control.

\subsection{Problem Formulation}
Typically, multi-agent cooperation can be mathematically modeled as multi-agent Markov games, which extends MDP to multi-agent setting \cite{littman1994markov}. A Markov game is defined as a five-tuple $(I, S, A, r, T)$, where $I$ is the set of $N$ agents, $S$ is the set of states, $A=A_{1} \times A_{2} \ldots \times A_{N}$ is the set of actions of all agents, and $r=\left(r_{1}, r_{2}, \ldots, r_{N}\right)$ where $r_i: S \times A \times S \rightarrow \mathbb{R}$ is the reward function of agent $i$. In the fully cooperative setting, we have $r_1=r_2,...,=r_N$. Given the current state and the actions of all agents, $T: S \times A \times S \rightarrow [0,1]$ is the probability distribution over the next states. Each agent $i$ aims to learn a policy $\pi_i : S_i \times A_i \rightarrow [0,1]$ to select the optimal action $a_i$ maximizing the accumulative rewards $R_{i}=\sum_{t=0}^{T} \gamma^{t} r_{i}^{t}$, where $\gamma$ is the is the discount factor and $T$ is the time horizon. We can formulate the joint policy of other agents as:
\begin{align*}
\pi^{-i}\left(a_{t}^{-i} \mid s_{t}\right)=\prod\nolimits_{j \in\{-i\}} \pi^{j}\left(a_{t}^{j} \mid s_{t}\right)
\end{align*}
Therefore, the objective of each agent $i$ is defined as follows:
\begin{align*}
\begin{split}
    & \max _{\pi^{i}} \eta_{i}\left[\pi^{i}, \pi^{-i}\right]
    \\ = &\mathbb{E}_{\left(s_{t}, a_{t}^{i}, a_{t}^{-i}\right) \sim T, \pi^{i}, \pi^{-i}}\left[\sum\nolimits_{t=1}^{\infty} \gamma^{t} R^{i}\left(s_{t}, a_{t}^{i}, a_{t}^{-i}\right)\right]
\end{split}
\end{align*}

Most existing multi-agent reinforcement learning methods assume that each agent can access all the other agents' policies during the training. In this paper, we consider a general scenario in a distributed manner, in which each agent $i$ has no knowledge of other agents' policies, and can only observe the historical states and actions of other agents, i.e., $\left\{s_{1: t-1}, a_{1: t-1}^{-i}\right\}$. Such a setting is also found in other decentralized MARL works \cite{zhang2018fully}\cite{qu2019value}.

\subsection{Hierarchical Decision-making Model for Multi-agent Cooperation}

\begin{figure}[t]
\centering
\includegraphics[width=.85\linewidth]{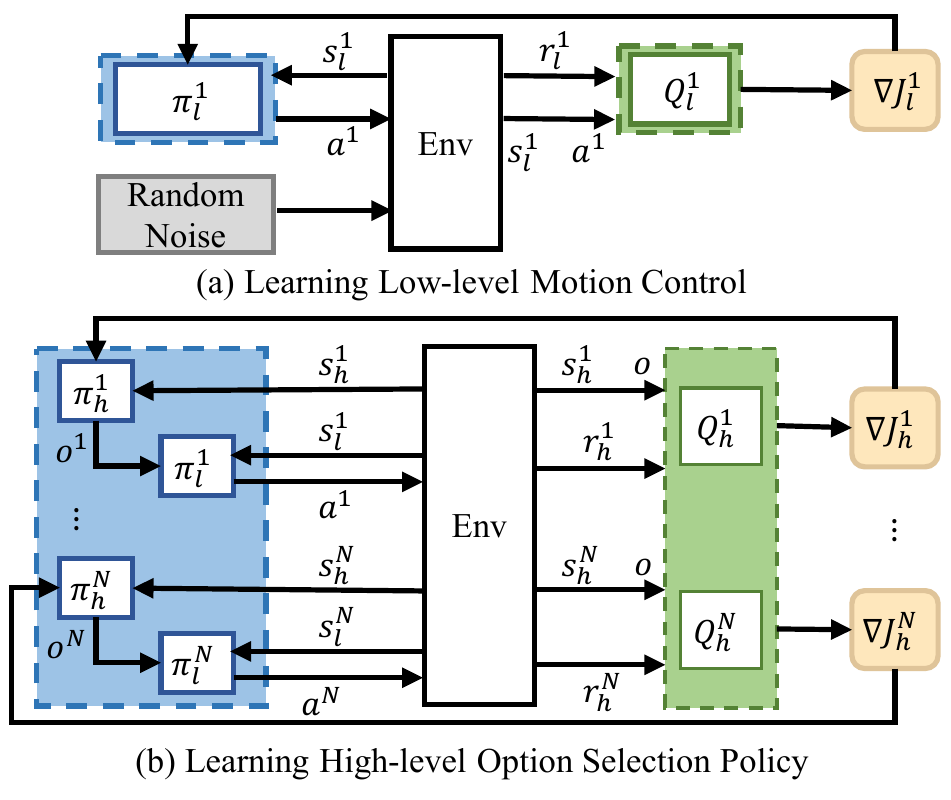}
\caption{Two-stage training structure of HERO. (a) Each individual agent learns different individual control policy with random noise in the first stage. (b) Multiple agents learn to select options in the second stage.}\label{fig2}
\end{figure}

\textbf{Hierarchical Policy Representation.} In this paper, we introduce a hierarchical reinforcement learning model for multi-agent cooperation, which decomposes an overall cooperation task into a hierarchy of discrete sub-tasks. The sub-tasks are also regarded as options, skills, or temporal abstractions in previous studies. For simplicity, we use the notation of option in the rest of this paper. In a two-layer hierarchical reinforcement learning model, the policy space of each agent is defined as follows:
\begin{align*}
    \pi^i=\left[\pi^i_h,\pi^i_l\right]
\end{align*}
where $\pi_h$ is the high-level policy to select the options and $\pi_l$ is the low-level policy to select primitive actions to perform the selected option.

A high-level option is defined as a three-tuple $o=(I_o,\pi_h,\beta_o)$, where ${I_o} \in \mathcal{S}_h$ is the initiation set, $\pi_{h}$: the option selection policy, and $\beta_o: \mathcal{\delta} \rightarrow[0,1]$ specifies the termination condition while executing $o$.

In the definition of high-level option, $S_h$ denotes the state space, and $A_h$ denotes the action space of high-level policy. The option $o$ can also be interpreted as the high-level action with the state set $S_h$. In the context of multi-agent cooperation, each agent $i$ cooperatively selects the option based on the current state $s^i_h$ and the inferred opponent options $o^{-i}_h$, which is shown in Fig.~\ref{fig2}. 

\textbf{Asynchronous Option Termination.} There are two modes for option termination: synchronous and asynchronous. Synchronous termination requires all agents to interrupt the existing option execution and select the next option synchronously, which is not feasible for fully distributed systems. Thus, we consider the asynchronous termination mode for each agent. At each time step $t$, the high-level layer of each agent $i$ will verify whether the current state $s^i_{h,t}$ satisfies the termination condition. In particular, if $\beta^i_o$ is equal to $1$, the existing option will terminate, and the high-level policy will select another option to execute until the overall task is finished.

\textbf{Hierarchical Reward.} Also, the reward of each agent $i$ can be divided into two parts for efficient training:
\begin{align*}
    r^i=\left[r^i_h, r^i_l\right]
\end{align*}
where $r^i_h$ and $r^i_l$ are the high-level and low-level rewards concerning team performance and individual control, respectively.

\subsection{Learning High-level Option Selection with Opponent Modeling}

\begin{figure}[t]
    \centering
    \includegraphics[width=.55\linewidth]{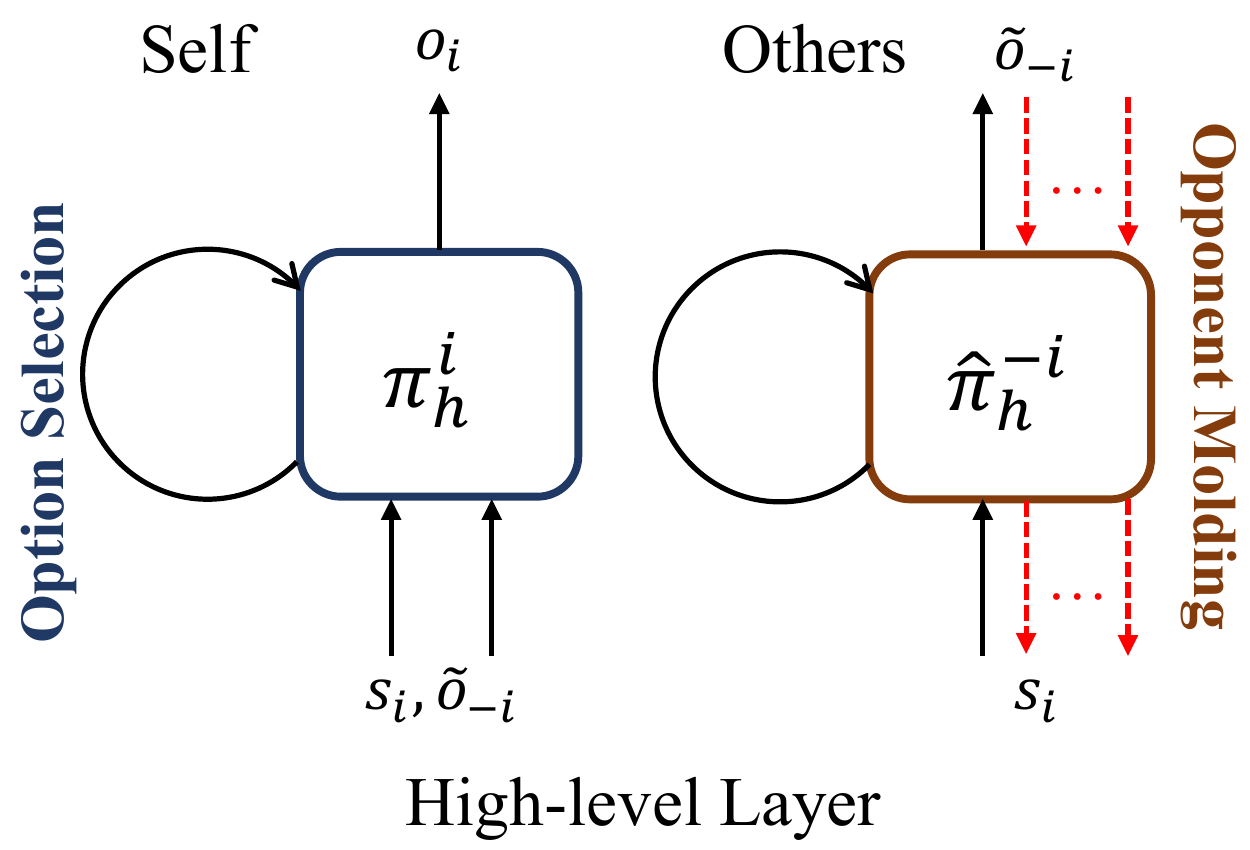}
    \caption{Illustration of the high-level opponent modeling in high-level layer. Each agent maintain a self policy network for its option selection and a opponent modeling network for other agents' option prediction.}
    \label{fig3}
\end{figure}

\textbf{Option-value Function.} The value function of each agent's option selection can be defined as follows:
\begin{align*}
    Q^i_h\left(s^i_{h,t},o_t  \right) = r^i_{h,t:t+c}+\gamma Q^i_h\left(s^i_{t+c}, o_{t+c}\right)-Q^i_h\left(s^i_{h,t}, o_t \right)
\end{align*}
where $\mathbf{o_t}=[o^i_t, o^{-i}_t]$ denotes the option selected by agent $i$ and the options selected by other agents. Here, we assume that each agent can observe all the previous option selections of others agents, i.e., $\left\{ o^{-i}_{1: t-1}\right\}$, at each time step. Besides, $r^i_{h,t}$ is the high-level reward received from environment which indicates the progress of the overall task and $r^i_{h,t:t+c}=\sum^{t+c}_{t} r^i_{h,t}$ denotes the accumulated high-level reward when performs the option $o^i_t$. Thereafter, the high-level transition that will be stored in the replay buffer $\mathcal{D}^i_h$ is expressed as follows:
\begin{align*}
\{(s^i_{h,t+k}, o^i_{t+k},o^{-i}_{t+k}, r^i_{h,t+k},s^i_{h,t+k+1})\}_{k=0}^{k=c}
\end{align*}

\begin{algorithm}[t]
\caption{Training High-level Cooperative Strategy with Opponent Modeling}
\begin{algorithmic}[1]
\STATE initialize the experience replay buffer $\mathcal{D}^i_{h}$ and the parameters {$\theta^Q_{i,h}$, $\theta^{\pi}_{i,h}$} for each agent i
\STATE initialize the experience replay buffer $\mathcal{D}^{-i}_{h}$ and the parameters $\pi^{-i}_{h}$ for opponent modeling
\FOR{episode $\gets 1$ \textbf{to} $M$}
    \STATE t $\leftarrow$ 0
    \STATE reset the environment and receive an initial states $s^i_{h,t}$ and $s^i_{l,t}$ for each agent $i$
    \STATE for each agent $i$ select high-level action (option) from the actor network $\pi^i_{h}$
    \STATE $o^i_t \leftarrow \pi^i_{h}$
    \WHILE{the task is not completed}
        \STATE pass the high-level option $o^i_t$ to low-level controller and generate low-level action $a^i_t$
        \STATE execute action $a^i_t$ and receive next state $s^i_{h,t+1}$
        \STATE compute the high-level team reward $r^i_h$
        \FOR{agent $i \gets 1$ \textbf{to} $N$}
            \IF{$s^i_{h,t+1}$ is a terminal state for option $o^i_{t}$} 
                \STATE choose next option $o^i_{t+1}$ from actor network $\pi^i_{h}$
                 \STATE store transition ($s^i_{h,t}, o^i_{t}, r^i_{h}, s^i_{h,t+1}$) in $\mathcal{D}^{i}_{h}$
                \STATE $o^i_{t} \leftarrow o^i_{t+1} $, $s^i_{h,t} \leftarrow s^i_{h,t+1}$ 
            \ELSE
            \STATE $o^i_{t} \leftarrow o^i_{t+1}$
            \ENDIF
            \STATE update the high-level critic network
            \STATE update the high-level actor network
        \ENDFOR
        \STATE store transition $(s^i_t, o^{-i}_t)$ in $\mathcal{D}^{-i}_{h}$
        \STATE t $\leftarrow t+1 $, $s_t \leftarrow s_{t+1} $
        \STATE update the opponent modeling networks
    \ENDWHILE
    \STATE update the target network parameters
\ENDFOR

\end{algorithmic}
\end{algorithm}

\begin{algorithm}[t]
\caption{Training Low-level Driving Skills with Intrinsic Reward Functions}
\begin{algorithmic}[1]
\STATE initialize the experience replay buffer $\mathcal{D}_{l}$ and the parameters {$\theta^l_Q$, $\theta^l_{\mu}$} for low-level controller
\FOR{episode $\gets 1$ \textbf{to} $M$}
    \STATE reset the environment and receive an initial state $s_0$
    \STATE t $\leftarrow$ 0
    \WHILE{$s_{t}$ is not a terminal state}
        \STATE select action from low-level actor network 
        \STATE $a_t \leftarrow \pi_l(\cdot \mid \mathbf{s}_{t})$
        \STATE execute action $a_t$ and receive next state $s_{t+1}$ and intrinsic reward $r_l$
        \STATE store transition $(s_t, a_t, r_t, s_{t+1})$ in $\mathcal{D}_{l}$
        \STATE t $\leftarrow t+1 $
        \STATE $s_t \leftarrow s_{t+1} $
        \STATE update the low-level critic network
        \STATE update the low-level actor network
    \ENDWHILE
    \STATE update the target network parameters
\ENDFOR

\end{algorithmic}
\end{algorithm}

\textbf{Agent Learning.} We propose a \textit{decentralized actor-critic} method for high-level policy learning. Each agent trains a decentralized critic network $Q^i_h\left(s^i_{h},o^i, o^{-i}; \theta^Q_{i,h} \right)$ to approximate the option-value function, while maintaining an actor network $\pi^i_h(s^i_h, o^{-i};\theta^{\pi}_{i,h})$ for option generation. Specifically, the critic network is trained by minimizing the loss:
\begin{align*}
    \mathcal{L}\left(\theta^Q_{i,h}\right)=\mathbb{E}_{{s^i_{h}}, o^i,o^{-i}, r^i_{h}\sim \mathcal{D}^i_h}\left[\left(Q^{i}_{h,t}\left({s^i_{h,t}}, o^{i}_t, o^{-i}_t\right)-y^i_{h}\right)^{2}\right]
\end{align*}
\begin{align*}
    y^i_{h}=r^{i}_{h,t:t+c}+\gamma Q^{i}_{h,t+c}\left(s^{i},o^i_{t+c},\hat{o}^{-i}_{t+c}) \right)
\end{align*}
where $\mathcal{D}^i_h$ is the experience replay buffer for offline high-level policy training. Then, the actor network can be trained with gradient ascent where the gradient is computed as:
\begin{align*}
\begin{split}
    &\nabla_{\theta^{\pi}_{i,h}} J\left(\theta^{\pi}_{i,h}\right)=\\
    &\mathbb{E}_{s^i_h \sim p^{\pi} , o^{i} \sim \pi^{i}_h}\left[\nabla_{\theta_{i}} \log \pi^{i}_h\left(o^{i} \mid s^{i}_h, \hat{o}^{-i}\right) Q^{i}_{h}\left(s^i_h, o^i, o^{-i}\right)\right]
\end{split}
\end{align*}
where $p^{\pi}$ denotes the state transition under all agents' policies $\pi={\pi_1,...,\pi_n}$ and $\hat{o}^{-i}$ denotes the inferred options of other agents generated from the opponent modeling network $\hat{\pi}^{-i}_h$.

\textbf{Opponent-modeling in Option Selection.} Opponent modeling is a feasible solution to the non-stationary issue in distributed reinforcement learning. Instead of modeling other agents' primitive actions, we propose to model their option selections reflecting their temporal abstraction during a few time steps. Then, two questions arise: \textit{how to use the opponent model} and \textit{how to train the opponent model}. 

For the model usage, we use the opponent model in the individual option selection to encourage the coordination behaviors. Each agent $i$ makes a decision based on both the current state $s^i_h$ and the inferred opponent options $\hat{o}^{-i}$, which is shown in Fig.~\ref{fig3}. Additionally, we use the latest opponent model to update the TD-target $y^i_h$:
\begin{align*}
    y^i_{h}=r^{i}_{h,t:t+c}+\gamma Q^{i}_{h,t+c}\left(s^i_h,\pi^i_h(s^i_{h,t+c}),\pi^{-i}_h(s^i_{h,t+c})) \right)
\end{align*}

Note that we input the option log probabilities of other agents directly into $Q^i_{h,t+c}$, rather than sampling.

For the opponent model training, we use a deep neural network to approximate the option selection policies of other agents and train the network by maximizing the log probability from the recent observation histories:
\begin{align*}
    \mathcal{L}\left(\theta^{-i}_{\pi,h}\right)=-\mathbb{E}_{s^{i}_{h}, o^{-i}_h}\left[\log \pi^{-i}_{h}\left(o^{-i}_h \mid s^{i}_h\right)+\lambda H\left(\pi^{-i}_{h}\right)\right]
\end{align*}
where $\pi^{-i}_{h}\left(o^{-i} \mid s^{i}_h\right)$ is the predicted probability of selected action $o^{-i}$ given the observation of agent $i$ and $H\left(\pi^{-i}_{h}\right)$ is the entropy of the policy distribution which is used to solve the over-fitting problem in deep learning.

\subsection{Learning Low-level Policies with Intrinsic Reward Functions}
Given the option, the objective of the low-level layer is to perform the option by selecting optimal primitive actions. The low-level individual control policy $\pi^i_l$ is represented by an individual deep neural network $\theta^i_l$ and trained with independent and parallel deep reinforcement learning. To successfully learn to perform the option, the intrinsic reward function $r^i_l$ is used for efficient training. In this paper, we adopt the \textit{soft actor-critic} method. According to the maximum entropy reinforcement learning theorem \cite{haarnoja2018soft}, the objective function of low-level policy is defined as follow:
\begin{align*}
    J(\pi^i_l)=\sum_{t=0}^{c} \mathbb{E}_{\left(s^i_l, a^i\right) }\left[r^i_l\left(s^i_l, a^i \right)+\alpha \mathcal{H}\left(\pi^i_l\left(\cdot \mid s^i_l\right)\right)\right]
\end{align*}
where $\mathcal{H}$ is the regulation term to encourage exploration and learn diverse policies to perform the specific option and $\alpha$ is the hyperparameter denote the weight of action exploration. Consequently, the gradient of each agent is:
\begin{align*}
\begin{split}
   & \nabla_{\theta^{i}_l} J\left(\theta^{i}_l\right)= \mathbb{E}\left[\nabla_{\theta^{i}_l} \epsilon \left[-\alpha \epsilon +Q^{i}_{l}\left({s}^i_l, {a}^i;o^i_h\right)\right]\right] \\
   & \text{where } \epsilon = \log \boldsymbol{\pi}^{i}_l\left(a^{i} \mid s^{i}_l\right)
\end{split}
\end{align*}

In practice, the training of critic can be realized by parameter sharing among distributed agents. Then, the critic network $Q^{i}_{l}\left({s^i_l}, {a^i};\theta^i_l, o^i_h\right)$ can be trained with deep Q-learning and the loss function is:
\begin{align*}
    \mathcal{L}(\theta^i_l)=\mathbb{E}_{s^i_l, a^i, r^i_l, s^{i,'}_{l}\sim \mathcal{D}^i_l }\left[\left(Q^i_l\left(s^i_l,a^i;\theta^i_l, o^i_h\right)-y^i_{l}\right)^{2}\right]
\end{align*}
\begin{align*}
    y^i_{l}=r^{i}_l+\gamma Q^{i}_l\left(s^{i,'},a^{'}_i; \theta^i_l;o^i_h \right)
\end{align*}
where $D^i_l$ is the experience reply buffer for off-policy training and $r^i_l$ is the intrinsic reward designed for the option learning.

\begin{figure}[t]
\centering \includegraphics[width=.65\linewidth]{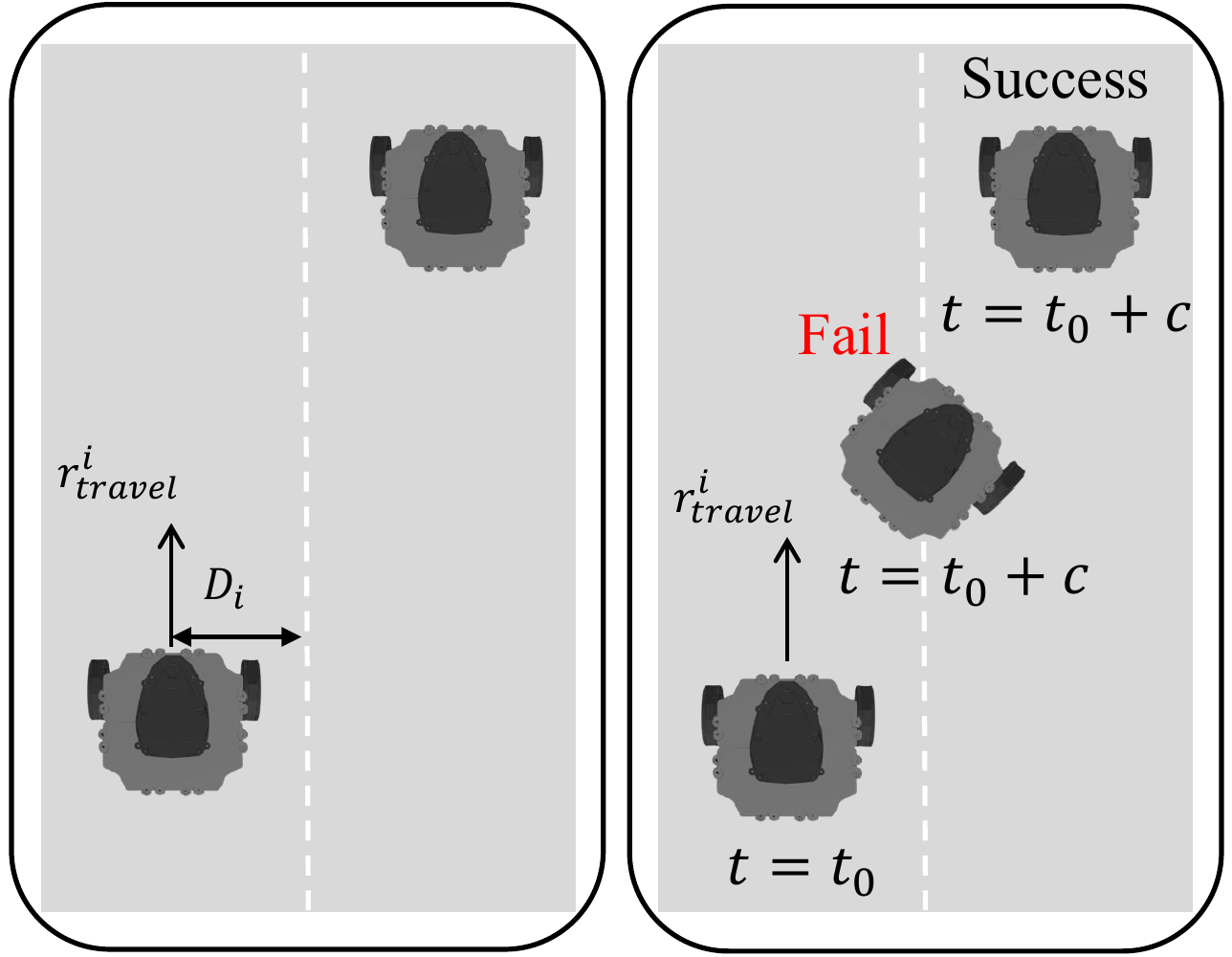}
\centering \caption{The left side of this figure shows the reward calculation for the training of driving in lane, while the right side shows the reward calculation for training of lane change.}
\label{Fig:reward_calculation}
\end{figure}

\section{Case Study: HERO for Multi-vehicle Cooperative Lane Change}\label{sec:case-study}
In this section, we present a case study of HERO on cooperative lane change to demonstrate the practicability of HERO in real-world applications.

\subsection{Cooperative Lane Change}
\begin{figure*}[t]
	\centering	\includegraphics[width=0.8\textwidth]{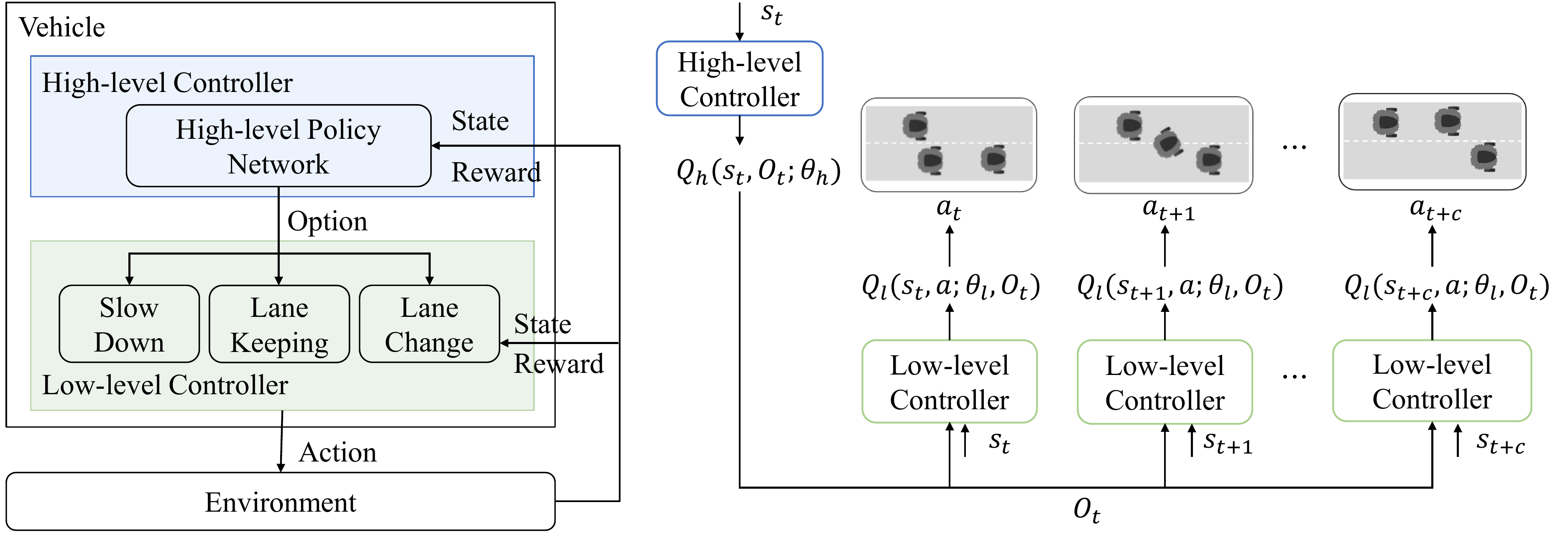}
	\caption{Hierarchical decision-making model for each vehicle.}
	\label{hrl_for_cooperative_lanechange_procedure} 
\end{figure*}

Lane change is one of the most essential and fundamental behaviors in autonomous driving, which involves the interaction of multiple vehicles in different lanes. Successful lane changes require drivers to account for several safety-related factors, including the road geometries, positions of ego vehicles, and cooperation with other vehicles. Inaccurate lane change leads to car accidents, congestion, and waste of energy. This paper aims to solve decision-making problems under different lane change scenarios. Many existing works only consider the single-agent driving scenario, while this work aims to leverage the cooperativeness of multiple vehicles to improve road safety and efficiency.

In this paper, we propose a two-layer hierarchical decision-making model, which includes a high-level controller and a low-level controller for cooperative driving. As shown in Fig.~\ref{hrl_for_cooperative_lanechange_procedure}, we first decompose the driving task into several sub-tasks, i.e., lane change, lane-keeping, and slow down. Each sub-task can be regarded as an option that the low-level controller can perform. It takes several time steps for the low-level controller to finish the option successfully. The objective of the high-level controller is to learn to select different options according to the current state that can maximize the traffic throughput while avoiding collisions. Specifically, the high-level controller's state, option, and reward function are defined in the following subsection.

\subsection{High-level State, Option and Reward Function Design}
\textbf{High-level State Space.} Here, we separate the state space of each vehicle into \textbf{$s_h^i$} and \textbf{$s_l^i$}, which are the state for high-level controller and low-level controller, respectively. The high-level state $s_h^i$ is defined as follows:
\begin{align*}
    s_h^i=\left[s_{lidar}^i,s_{speed}^i,s_{laneID}^i\right]
\end{align*}
where $s_{lidar}^i$ is the raw observation from the lidar sensor, which denotes the distance with other vehicles from $360$ degrees, $s_{speed}^i$ is the vehicle's speed, and $s_{laneID}^i$ is the vehicle's current lane identifier.

\textbf{High-level Option Space.} The action space of high-level controller is to choose \textit{which option to execute}. Here, we consider the discrete action space defined as follow:
\begin{align*}
    A_h^i=\left[\text{keep lane}, \text{slow down}, \text{accelerate}, \text{lane change}\right]
\end{align*}

We define four types of high-level options according to human driving behaviors, where each option corresponds to a learned low-level control policy.

\textbf{High-level Team Reward.} The design of high-level team reward is related to the overall objective. Typically, safety and efficiency are two main concerns in cooperative merge scenarios. In terms of safety, each vehicle should not only avoid collision with the front vehicle when driving in the current lane but also pay attention to the vehicles in the merged lane. Once the collision happens, we will assign a negative reward $r_{col}$ to the vehicles as the penalty, and the current training episode will be ended and restarted. To avoid traffic congestion, we encourage the vehicle to run as fast as possible by giving a positive reward $r_{travel}$ to vehicles. Hence, the total reward of each vehicle is designed as follows:
\begin{align*}
    r_{h}^i=\alpha r_{col}+(1-\alpha)r_{travel}^i
\end{align*}
where $r_{col}$ is the penalty of collision, $r_{travel}$ is the reward for moving forward, and $\alpha$ is a hyperparameter to control the weight of collision avoidance and move forward.

\subsection{Low-level State, Action and Intrinsic Reward Function Design}
\textbf{Low-level State Space.} Previous works rely on the given waypoint data of the vehicles and use PID controller to drive. In this paper, we consider learning driving skills based on vision, which is similar to how human drive. Therefore the low-level state space $s_l^i$ is defined as follows:
\begin{align*}
    s^i_l=\left[s^i_{img}, s^i_{speed}, s^i_{laneID}\right]
\end{align*}
where $s^i_{img}$ denotes the captured image, $s^i_{speed}$ represents the speed, and $s^i_{laneID}$ is the identifier of current lane.


\textbf{Low-level Action Space.} Because the low-level controller directly determines the motion of the vehicle, the action space of the low-level controller is the continuous linear speed and angular speed. If the high-level option selection is lane-keeping, then the linear and angular speeds will remain the same compared to the last step. Otherwise, the action spaces of different skills are defined as follows:
\begin{align*}
    A^i_{\text {linear}}= \begin{cases}0.04:0.08 & \text { slow down } \\ 0.08:0.14 & \text { accelerate } \\ 0.1:0.2 & \text { lane change }\end{cases}
\end{align*}
\begin{align*}
    A^i_{\text {angular}}= \begin{cases}-0.1:0.1 & \text { slow down } \\ -0.1:0.1 & \text { accelerate } \\ 0.12:0.25 & \text { lane change }\end{cases}
\end{align*}
where $A^i_{linear}$ denotes the action space for linear speed and $A^i_{angular}$ is the action space for angular speed. To prevent unsafe actions and improve exploration efficiency, we set different lower and upper bounds of the linear speed and angular speed during the training.

\textbf{Low-level Intrinsic Reward Functions.} In contrast to the high-level team reward which directly related to the success of the overall task, the principle of \textit{intrinsic reward} is to provide feedback for learning different skills. Here, we categorize the intrinsic rewards into two types: $r^i_\text{driving in lane}$ and $r^i_\text{lane change}$. An illustration of reward calculation is shown in Fig.~\ref{Fig:reward_calculation}. In particular, $r^i_\text{driving in lane}$ is used for $o^i_h=\left[\text{slow down}, \text{accelerate}, \text{keep lane}\right]$ calculated as follows:
\begin{align*}
    r^i_{\text {driving in lane}}=\beta \cdot r^i_{deviate} + (1-\beta) \cdot r^i_{travel}
\end{align*}
where $r^i_{deviate}$ denotes the distance deviated from the center of current lane, $r^i_{travel}$ is the travel distance used to encourage the movement of each vehicle, and $\beta$ is a hyper parameter controlling the weights of deviated distance and travel distance.

Similarly, the intrinsic reward for lane change skill is designed as follows:
\begin{align*}
    r^i_{\text {lane change}}= \begin{cases}20 & \text { success } \\ -20 & \text { fail } \\ r^i_{travel} & \text { else } \end{cases}
\end{align*}

For the lane change skill, we assign a positive reward of $20$ when the vehicle successfully changes the lane and assign a negative penalty if the vehicle fails to change lane within a predefined maximum time step.

\begin{figure}[t]
\centering
\includegraphics[width=\linewidth]{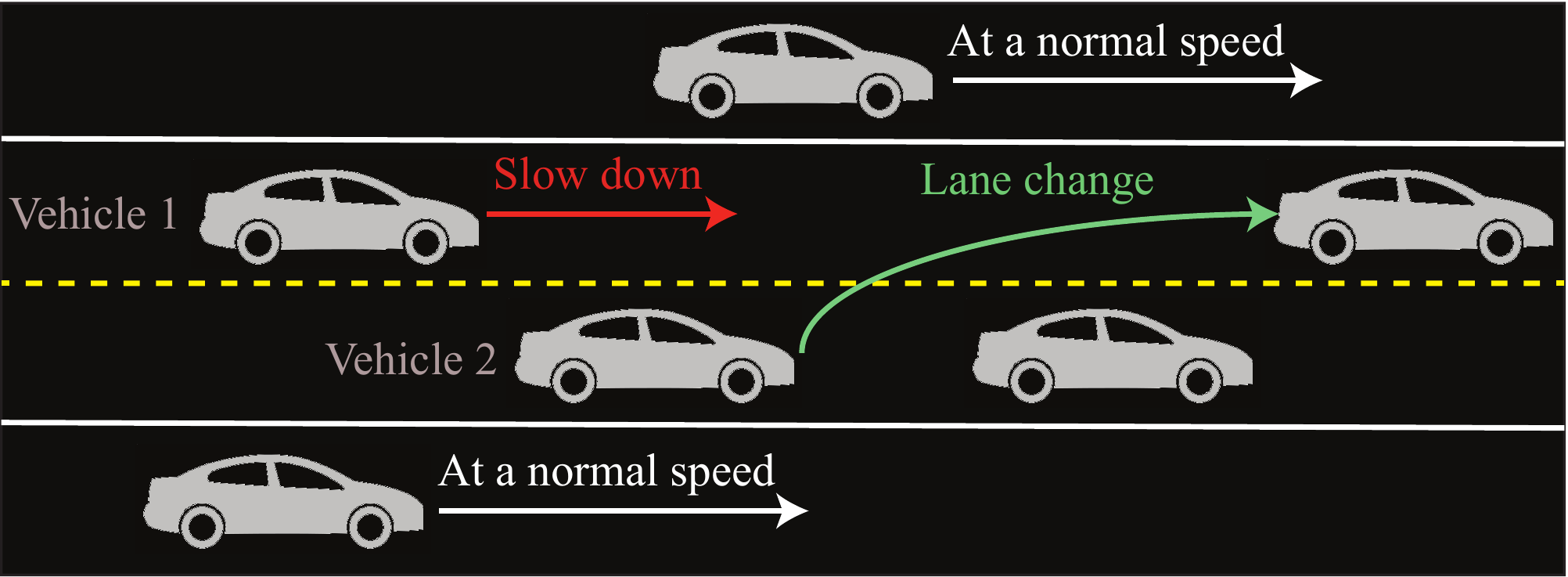}
\caption{Illustration of the cooperative lane change scenario, where the vehicle $1$ should coordinate with vehicle $2$ to avoid the collision when vehicle $2$ is performing lane change}
\label{Fig:lanechange}
\end{figure}

\section{Performance Evaluation}\label{sec:exp}
This section presents the results of applying our approach in cooperative lane change scenarios, which is shown in Fig.~\ref{Fig:lanechange}. We first create a simulation environment for training and then build a real-world testbed for further evaluation. We compare the performance of our approach with several well-known reinforcement learning approaches. 

\begin{table}[t]
\centering
\caption{Hyperparameters for Training}
\normalsize
\begin{tabular}{|c|c|}
\hline 
\textbf{Hyperparameter} & \textbf{Value}\\
\hline 
Training episode & $14,000$\\
\hline
Episode length & $30$\\
\hline
Buffer capacity & $100,000$\\
\hline
Batch size & $1024$\\
\hline
Learning rate & $0.01$\\
\hline
Discount factor $\gamma$ & $0.95$\\
\hline
Dimension of the hidden layer & $32$\\
\hline
Target network update rate & $0.01$\\
\hline
\end{tabular}
\label{tab:hyperparameter}
\end{table}

\begin{figure*}[t]
\centering
\includegraphics[width=\linewidth]{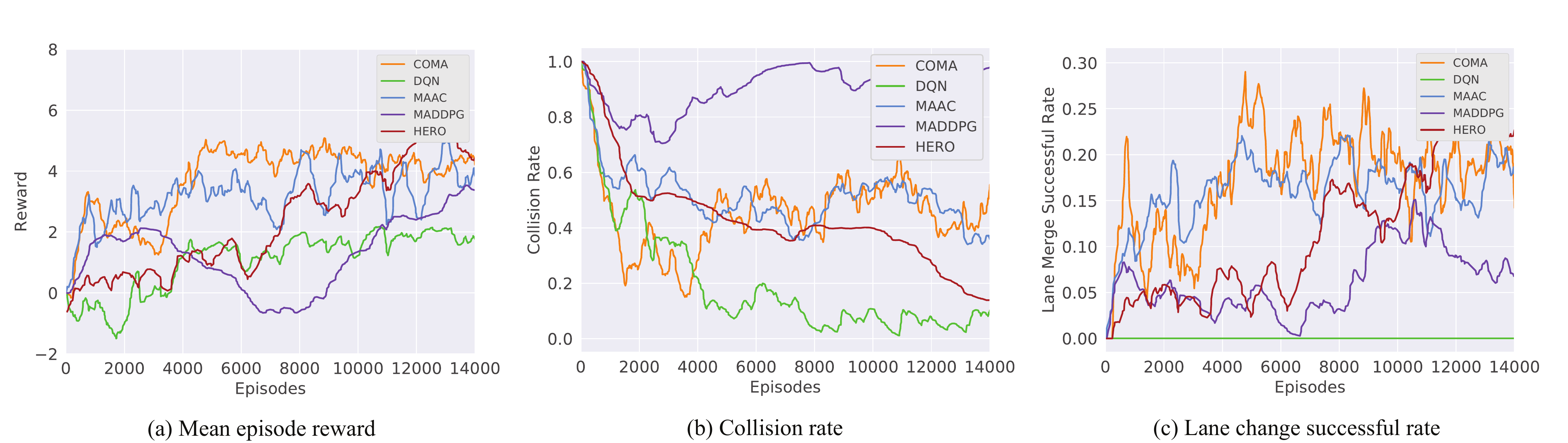}
\caption{Comparison of the learning curve of different approaches in the cooperative lane change scenarios.}
\label{fig:high-level-comparision}  
\end{figure*}

\subsection{Baseline Approaches}
In this paper, we apply different MARL approaches to cooperative lane change scenarios, including distributed approaches (DTDE) and centralized training with decentralized execution (CTDE) approaches. Specifically, the considered methods for comparison include:
\begin{itemize}
    \item \textit{Independent Deep Q-learning (DQN)}. It is a distributed learning approach, in which each agent trains a Q-network using its local observation and \textit{shared team reward}. Each agent applies the $\epsilon$-greedy strategy for action exploration during the training. 
    \item \textit{Counterfactual policy gradient (COMA)}. It is a standard CTDE approach where the centralized critic is trained with Q-learning. The actor network is trained with the counterfactual policy gradient theorem.
    \item \textit{Multi-agent mixed actor-critic (MADDPG)}. It is another approach adopting the framework of CTDE. Unlike COMA, it considers the environment with explicit communication and trains a centralized critic for each agent, allowing each agent to have different reward functions.
    \item \textit{Multi-Actor-Attention-Critic (MAAC)}. It is the state-of-the-art approach that extends the attention mechanism, which is widely used in image processing and natural language processing tasks, to multi-agent reinforcement learning. It trains an actor-attention-critic network for each agent and allows parameter sharing to improve the learning efficiency. Noted that, MAAC uses decentralized critics with a decentralized actor with parameter sharing.
\end{itemize}

\subsection{Training Environment and Evaluation Metrics}
To evaluate our approach, we first create a simulation environment based on Gazebo, a popular physical simulation engine in robotics \cite{koenig2004design}. The simulation environment includes the lane scenario and multiple vehicles equipped with cameras and lidar sensors. Besides, we provide several RL-friendly APIs that are convenient for algorithm implementation. Fig.~\ref{Fig:Gazebo} shows the simulation scenarios of the cooperative lane change. When the vehicle in front of vehicle $2$ stops or moves slowly, vehicle $2$ needs to control itself to perform lane change which needs the coordination among vehicle $1$ and vehicle $2$. We use a conventional neural network to encode the image data and a multi-layer fully-connected neural network as the critic network for policy implementation. The dimension of the hidden layer is set to $32$ in all the algorithms. 

At the beginning of each episode, we randomly initialize several positions of the vehicles and set the same initial speed to them. Besides, the episode length is set to $18$ time steps. The hyper-parameters of our network architecture and learning algorithms are presented in Tab.~\ref{tab:hyperparameter}. When the vehicles get collision, or the max episode length is reached, the program will call the \textit{reset} function provided by our testbed to reset the environment and continue the training.

In this paper, we consider four evaluation metrics to measure the performance of different approaches, including:
\begin{itemize}
    \item \textit{Mean episode reward}: the average reward of each time step when sampling the replay buffer from the different episodes.
    \item \textit{Collision rate}: the collision ratio of the vehicles in each episode.
    \item \textit{Lane merge successful rate}: the proportion of vehicles that successfully merge in each episode.
    \item \textit{Mean speed}: the average of the vehicle's speeds at each time step.
\end{itemize}

\begin{figure}[t]
\centering
\includegraphics[width=\linewidth]{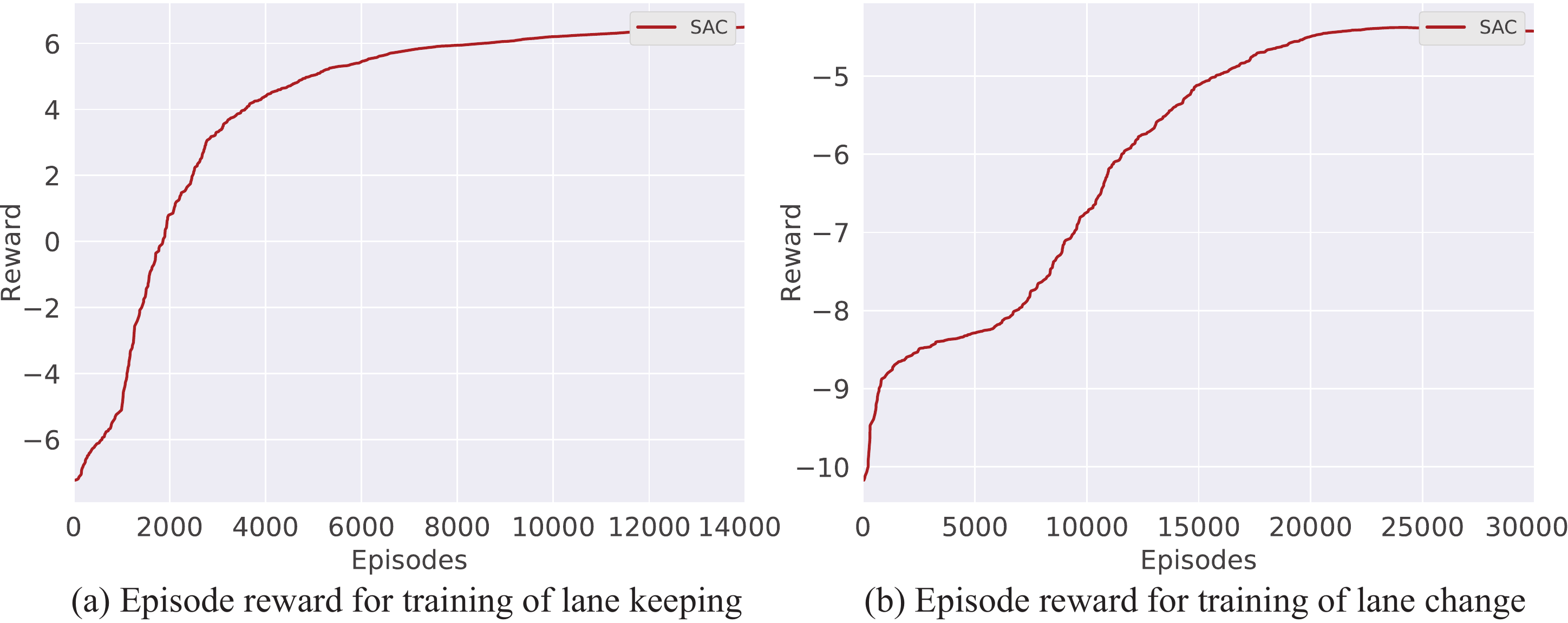}
\caption{Episode reward of learning different low-level skills.}
\label{fig:low-level-reward}
\end{figure}

\subsection{Learning Low-level Skills}

\begin{figure}[t]
\centering
\includegraphics[width=.5\linewidth]{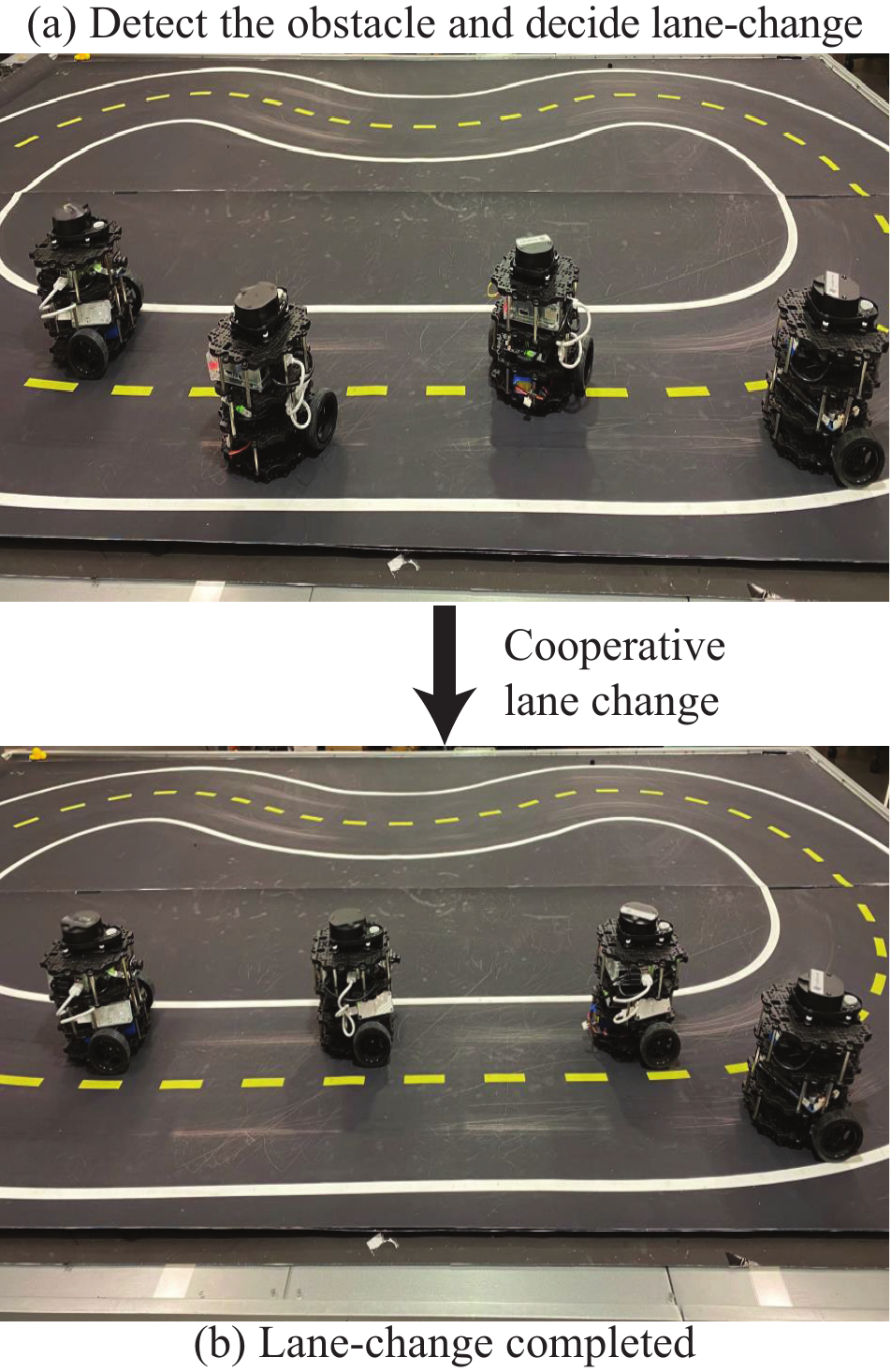}
\caption{Real-world evaluation}
\label{fig:real-world-eva}
\end{figure}

Before training cooperative strategies, we first train the low-level skills, including lane tracking and lane change. We create parallel training environments with different intrinsic reward functions so that the low-level learned skills can be further shared. To improve learning efficiency, we set a lower angular speed in training the lane-keeping skill and set a higher angular speed when training the lane change skill. Fig.~\ref{fig:low-level-reward} shows the episode rewards during the training process. The experimental results imply that the soft actor-critic algorithm can successfully converge in training the lane tracking and lane change tasks. To be noticed, the episode reward of learning lane change policy remains a low value before $5,000$ episodes. The reason is that the agent will explore the action space at the beginning to maximize the entropy of action probability. The exploration strategy makes the low-level skill model robust.

\subsection{Learning High-level Cooperative Policy}

\begin{figure}[t]
\centering
\includegraphics[width=\linewidth]{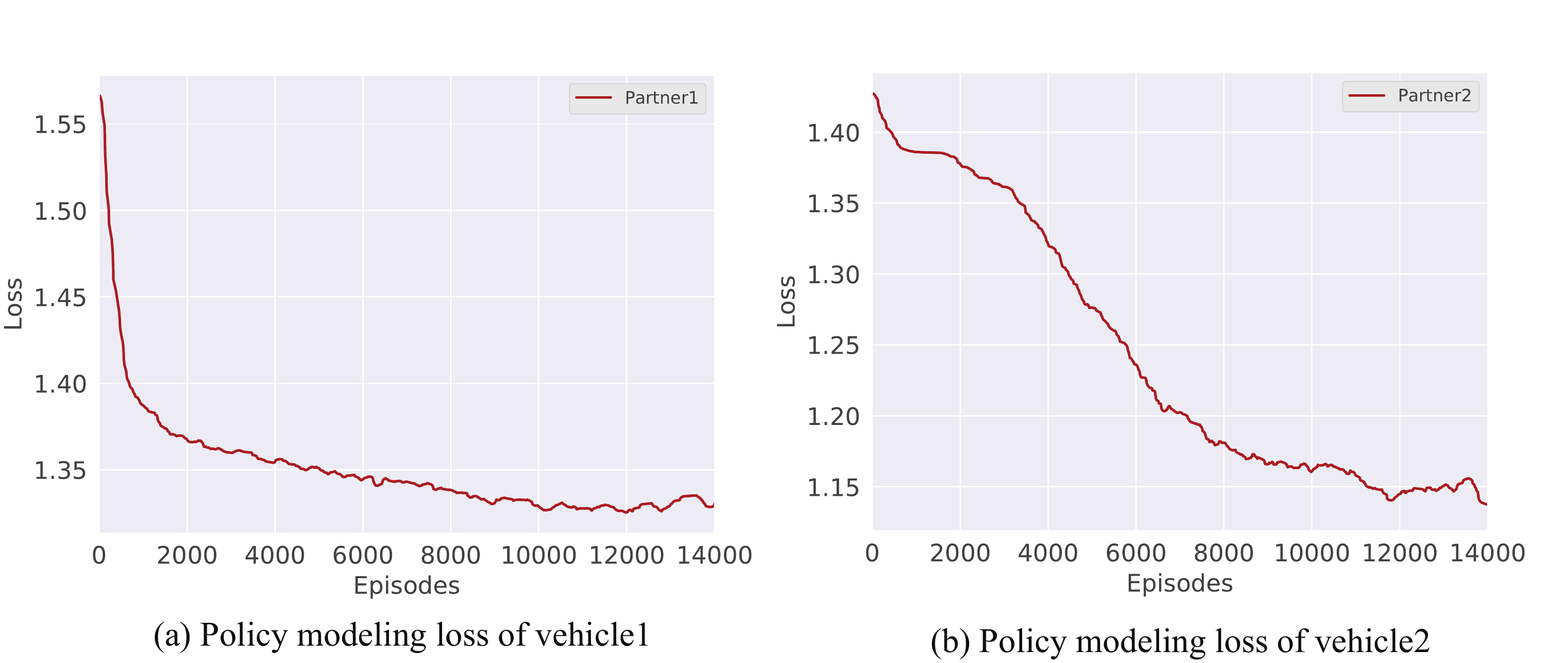}
\caption{Learning loss of modeling other partners' policies.}
\label{fig:modeling-loss}
\end{figure}

\begin{figure}[t]
\centering
\includegraphics[width=0.6\linewidth]{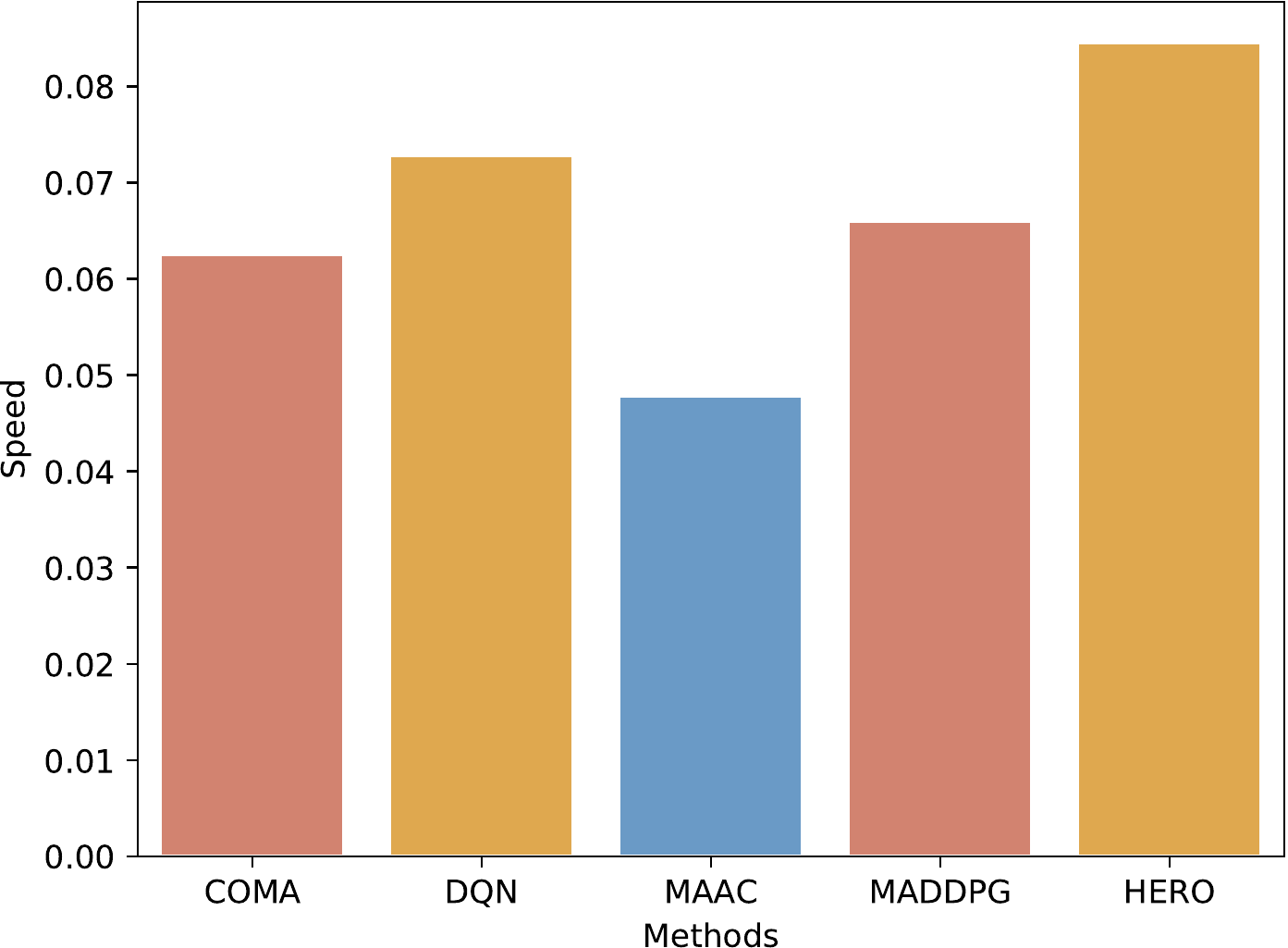}
\caption{Learning high-level cooperative policy in the simulation environment we created.}
\label{Fig:Mean Speed}
\end{figure}

\begin{figure}[t]
\centering
\includegraphics[width=.5\linewidth]{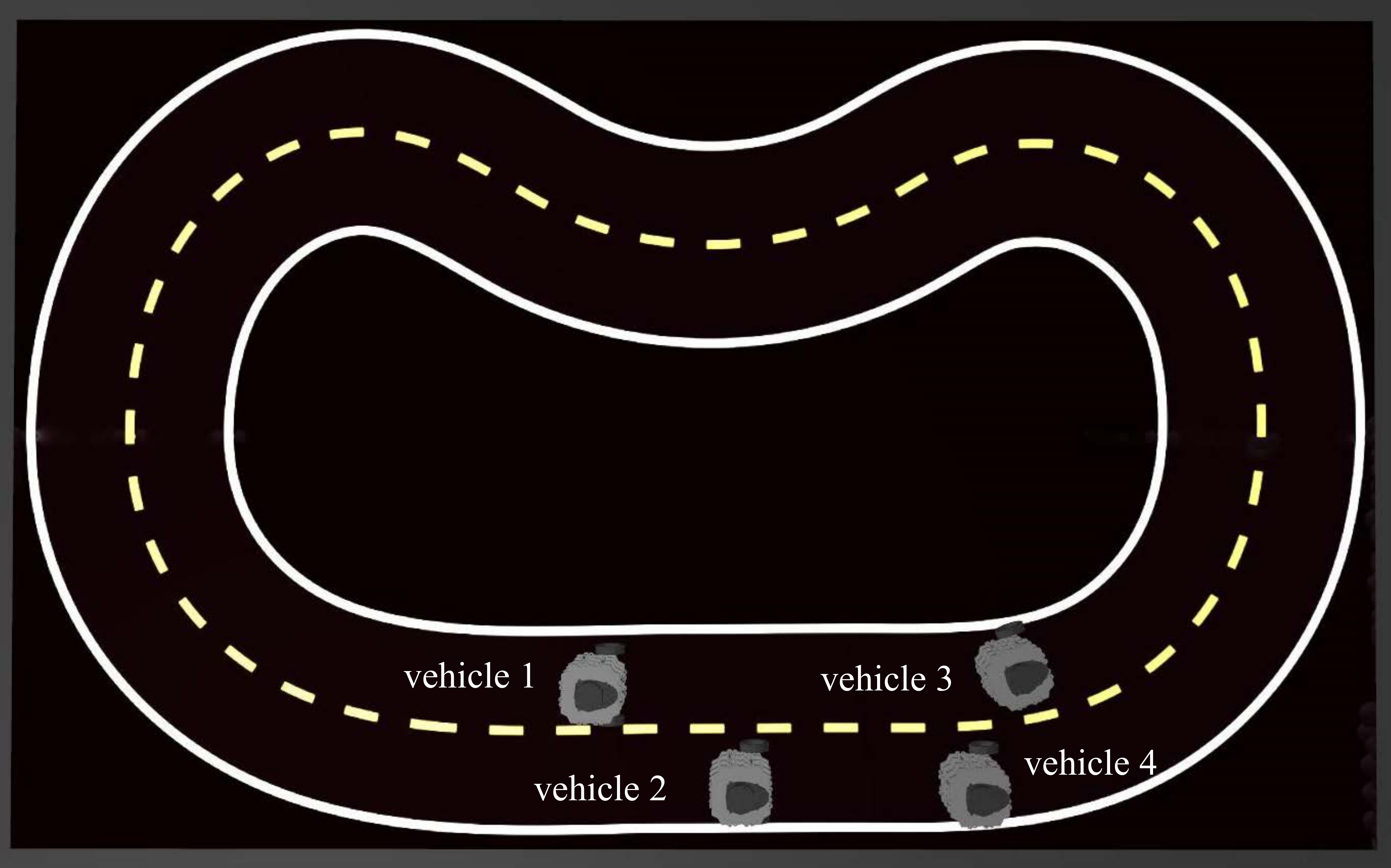}
\caption{Training in the Gazebo simulator.}
\label{Fig:Gazebo}
\end{figure}

After training the low-level skills in the single-vehicle environment, we gradually increase the number of vehicles to learn the coordination policy. As shown in Fig.~\ref{fig:real-world-eva}, we set up four vehicles in a double-lane track scenario. Vehicle $4$ is set with a plodding speed to simulate traffic congestion or traffic accident. The other vehicles are initialized with an average speed and random positions. Once the collision happens during lane-changing or lane-keeping, each vehicle will receive a negative reward of $-20$, and the episode will end and restart. 

During the training, each vehicle will train opponent models to predict others' actions and use the opponent model to stabilize the offline training of the Q-value network. Fig.~\ref{fig:modeling-loss} shows the loss of modeling other vehicles' policies from vehicle 2's perspective. As we can see, the vehicle $ 1$'s action prediction model become converged at a fast speed while the vehicle $ 3$' action prediction model become converged after 12000 episodes. The difference in convergence speed also illustrates the different interactions among vehicles. 

\begin{figure}[t]
\centering \includegraphics[width=0.9\linewidth]{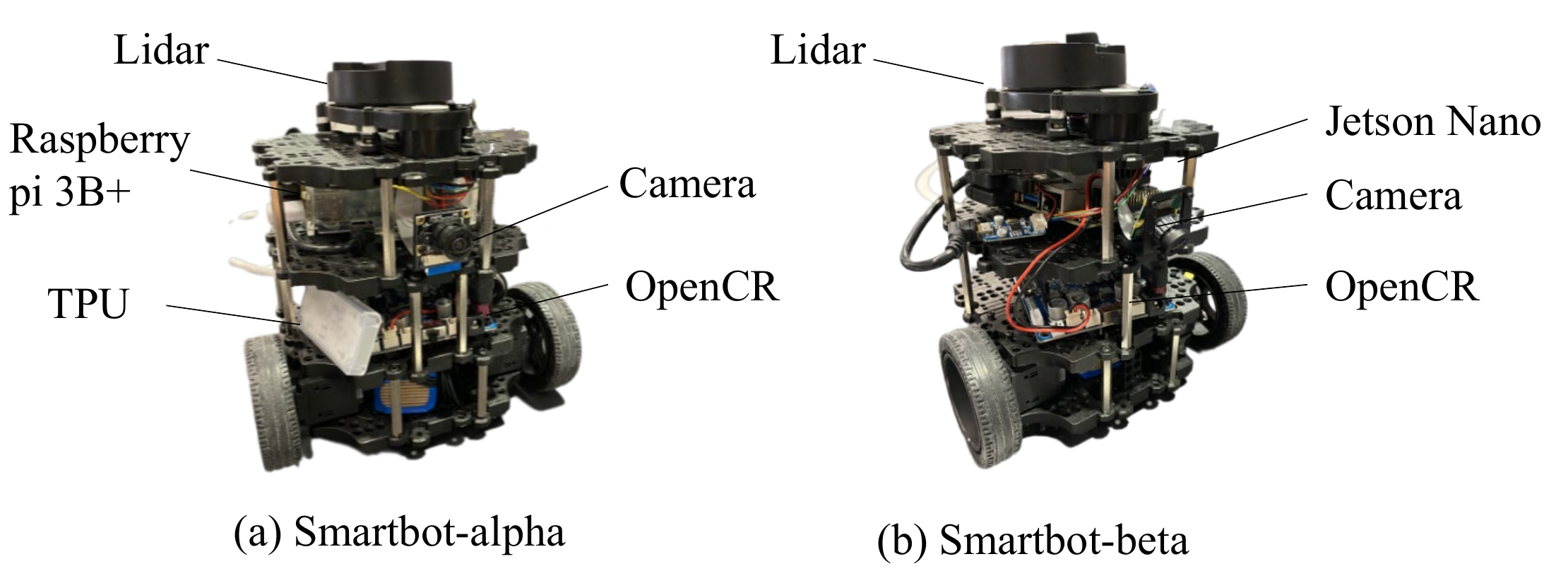}
\centering \caption{We develop two prototypes of vehicles in the real-world testbed. In this experiment, we use the first prototype in the testing.}
\label{fig:robts}
\end{figure}

\begin{table}[t]
\centering
\caption{Performance evaluation in real-world platform}
\begin{tabular}{|c|c|c|c|}
\hline
\diagbox{Method}{Metrics} & Collision Rate & \makecell[c]{Successful \\ Rate} & \makecell[c]{Mean \\ Speed} \\
\hline 
COMA & $0.35$ & $0.65$ & $0.06344$ \\
\hline 
Independent DQN & $1.0$ & $0.0$ & $0.05395$ \\
\hline
MAAC & $0.25$ & $0.65$ & $0.0625$ \\
\hline
MADDPG & $0.95$ & $0.5$ & $0.07029$ \\
\hline
Ours & $0.2$ & $0.8$ & $0.072$ \\
\hline
\end{tabular}
\label{tab:real-performance}
\end{table}

Fig.~\ref{fig:high-level-comparision} shows the performance of different RL approaches in terms of \textit{mean episode reward, collision rate} and \textit{lane change successful rate}. As shown in Fig.~\ref{fig:high-level-comparision}(a), our method achieved the highest episode reward than other baselines. Besides, the lower bound of our method is highest than others which also implicitly shows the stability of our approach. Fig.~\ref{fig:high-level-comparision}(b) and Fig.~\ref{fig:high-level-comparision}(c) shows the collision rate and lane change successful rate among different approaches. Almost all the RL approaches can reduce the collision rate after 14000 training episodes except MADDPG. We found that MADDPG has a lower learning efficiency as it still maintains a higher collision rate than others. Both DQN and our method reach the lowest collision rate. However, the lane change success rate is near $0$. We found that vehicle $2$ learn a policy to drive slowly instead of changing the lane during the training. Instead, the vehicle trained with our method can successfully perform cooperative lane change without collision with others. Moreover, Fig.~\ref{Fig:Mean Speed} shows that our method achieves the highest mean speed, which is $0.08$, and the vehicles trained with MAAC received the lowest speed of $0.048$. In general, our method shows advantages in terms of \textit{safety} and \textit{efficiency}.

\subsection{Real-world Experiments}

To investigate the gap between simulation and reality, we deploy the learned policies in simulation to a real-world testbed, which consists of a two-lane track and multiple vehicles shown in Fig.~\ref{fig:robts}. Each vehicle is equipped with a camera, lidar, and edge server. We run $20$ episodes for each MRRL method with several random initial positions. Besides, we run a master node on the server to monitor the state of each vehicle and calculate the collision rate, lane-merging successful rate, and mean speed.

As shown in Tab.~\ref{tab:real-performance}, our approach reaches a low collision rate and high speed than others, which is $0.2$ and $0.07$, respectively. Besides, the collision rate of MAAC is lower than other baselines such as COMA and MADDPG. However, the performance of Independent DQN is different in simulation. We argue that the diversity and robustness of DQN are poor than other policy-based approaches.

\section{Related Work}\label{sec:related-work}
This section summarizes the previous works related to multi-agent reinforcement learning and hierarchical reinforcement learning in Sec.~\ref{sec:marl} and Sec.~\ref{sec:hrl}, respectively.

\subsection{Multi-agent Reinforcement Learning}\label{sec:marl}
Recent works on reinforcement learning (RL) have shifted from single-agent reinforcement learning to multi-agent reinforcement learning (MARL). MARL corresponds to the learning problem in multi-agent systems in which multiple agents learn simultaneously in a shared environment. Existing MARL approaches can be classified into centralized RL, centralized training with decentralized execution (CTDE), and distributed RL. Centralized RL trains a centralized network mapping from the state space to all the agents' joint action space. Such an approach suffers from the high dimension issue of joint state and action spaces, making it hard to be extended to large-scale scenarios.

Some previous works adopt the paradigm of centralized training with decentralized execution (CTDE). Lowe et al. proposed a multi-agent actor-critic (MADDPG) approach and applied it for cooperative navigation \cite{lowe2017multi}. Foerster et al. proposed counterfactual policy gradient (COMA), using a counterfactual baseline to address the credit assignment problem in MARL \cite{foerster2017counterfactual}. Similar works of CTDE can also be found in flocking \cite{zhu2020multi} and pathfinding \cite{damani2021primal}. However, the number of features of the centralized critic network needs to be scaled up linearly (in the best case) or exponentially (in the worse case) with the increases of the number of agents \cite{lyu2021contrasting}. Besides, different agents may have different influences, and decentralized critics are expected to perform better in these scenarios.

Recent developments in DRL can be further divided into two types: independent RL and decentralized critics with decentralized actors. Independent RL is self-interested and does not consider the states and actions of other agents \cite{sivanathan2020decentralized}. Independent RL has good scalability but poor cooperation performance. An alternative method is to use decentralized critics with decentralized actors(DTDE). For example, Zhang et al. proposed a general decentralized MARL framework \cite{zhang2018fully} and Iqbal et al. proposed an actor-attention-critic approach for multi-agent cooperation \cite{iqbal2019actor}. However, these methods either rely on explicit information sharing or use end-to-end RL models, which are not adaptive to complex and complicated tasks. Moreover, these end-to-end RL models lack interpretability on the interaction among agents.

\subsection{Hierarchical Reinforcement Learning}\label{sec:hrl}
Hierarchical reinforcement learning (HRL) is a type of reinforcement learning (RL) that leverages the hierarchical structure of a given task learns a hierarchical policy \cite{dietterich1998maxq}\cite{jong2008hierarchical}. Dayan et al. developed the first HRL work and proposed feudal reinforcement learning (FRL), where the high-level \textit{managers} learn how to set the tasks to the \textit{sub-managers} while the sub-managers learn how to perform the sub-tasks \cite{dayan1994feudal}. In this way, a complex task can be divided into small sub-tasks that are easier to be solved \cite{dayan1994feudal}. Kulkarni et al. introduced \textit{intrinsic reward} design for low-level tasks and used a deep neural network to approximate the value function \cite{kulkarni2016hierarchical}. The proposed intrinsic reward mechanism motivates the agents to explore new behavior for its own sake.

Some previous works have applied hierarchical reinforcement learning for multi-agent cooperation. For example, Dietterich proposed the MAXQ framework to learn the hierarchical policy of each agent \cite{dietterich2000hierarchical}. Besides, Ghavamzadeh et al. proposed a COM-Cooperative HRL for multi-agent communication \cite{ghavamzadeh2004learning}. Recently, Ahilan et al. proposed a deep MAHRL approach, in which a central manager policy chooses subtasks for other workers simultaneously \cite{ahilan2019feudal}. Similar deep MAHRL approaches can also be found in \cite{chakravorty2019option}. However, all of these works either only consider the discrete action space of each agent or rely on centralized training.

\section{Conclusion and Future Directions} \label{sec:dis-con}
This paper studies how distributed multiple agents learn to cooperate in continuous action space and the non-stationarity issue. Unlike the traditional end-to-end RL methods, we propose a hierarchical reinforcement learning approach for distributed multi-agent cooperation. The agent cooperation is efficiently learned in high-level discrete action space, while the low-level individual control is handled by independent reinforcement learning. We incorporate \textit{opponent modeling} in the high-level layer to encourage cooperation and stabilize Q-learning. Then, we present a case study of cooperative lane change and conduct extensive experiments via simulation and a real-world testbed. In the future, we will investigate deeper hierarchy discovery and the theoretical guarantee. Currently, the design of task decomposition and state-action space in different layers still requires human knowledge. It is still an open question how to discover the hierarchical architecture automatically. Besides, most existing MARL approaches are trained and evaluated in simulation. The gap between simulation and reality also deserves more investigation.

\section{Acknowledgment}
The research is supported by Hong Kong RGC TRS (project T41-603), RIF (projects R5009-21 and R5060-19), CRF (projects C5026-18G and C5018-20GF), and GRF (project 15204921).

\bibliographystyle{IEEEtran}
\bibliography{main}

\end{document}